\documentstyle[preprint,eqsecnum,amssymb,aps,prb]{revtex}
\begin{document}
\draft
\preprint{}
\title{Dipolar-broadening of $\bbox{I}={\bbox{1}\over\bbox{2}}$-NMR
  spectra of solids}
\author{Jens Jensen}
\address{\O rsted Laboratory, Niels Bohr Institute,
Universitetsparken 5,
2100 Copenhagen, Denmark}
\date{\today}
\maketitle
\begin{abstract}The equations of motion of dipolar-coupled spins of
$I=1/2$ placed on a rigid lattice are solved
approximately in the high-temperature and high-field limit. The
NMR-spectra predicted by this theory are in close agreement
with both the theoretical spectral moments, up to the eighth in
the case of a simple cubic lattice, and the extremely accurate
experimental results which have been obtained in the case of CaF$_2$.
The theory is compared with the recent NMR-experiment of Lefmann {\it
  et al.}\ on $^{13}$C diamond. It predicts a double-peak
splitting of the NMR-spectrum when the field is applied along [111], in
accordance with the experiment, though the widths of the calculated
resonance lines are smaller than observed.
\end{abstract}


\section{Introduction}
\label{sec:level1}
Recently Lefmann {\it et al.}\cite{Lef} have measured the NMR-spectra
on single crystals of the new material $^{13}$C diamond. The nuclear
spin of the $^{13}$C isotope is $1/2$, and in the case of
a field applied along $[111]$, the square of the magnetic
dipole--dipole interaction
between two nearest-neighboring spins in the direction of the field is
an order of magnitude stronger than the interaction between any other
neighbors. This leads to a splitting of the resonance into two peaks,
a phenomenon which has only previously been observed in dense
systems of molecules,\cite{Pake,Ped,Par1} The $^{13}$C diamond
constitutes a system of $I=1/2$ nuclei interacting through
the dipole force between their magnetic moments and, because the
Debye temperature of the diamond crystal is high, the lattice may be
considered to be rigid. The same model applies for the NMR in CaF$_2$
except that the $I=1/2$ nuclei, the F-ions, are placed on
a simple cubic lattice.

The simplest problem of NMR in solids is the one where identical
nuclei of $I=1/2$ are placed on a rigid lattice. In this case no
quadrupole moment or quadrupolar broadening effects have to be
considered, and lattice vibrations are neglected. The calculation of
the dipolar broadening of the NMR is, nevertheless, a complicated
many-body problem with no small expansion parameter. Van Vleck
calculated the second and fourth moments of the spectral
distribution,\cite{VV} which indicate that the line shape is
neither Lorentzian nor Gaussian. The real-time response function
$F(t)$, the so-called free-induction-decay (fid) curve, derived from
experiments\cite{Bruce} on CaF$_2$, was found by Abragam\cite{Abra} to
be accurately described by the function
\begin{equation}
F(t)=e^{-a^2t^2/2}\sin(bt)/bt\,,\label{1}
\end{equation}
when the parameters $a$ and $b$ were adjusted so to get the right
second and fourth moments of the spectral function. In the early
seventies the calculation of the moments was extended  to include
the sixth and eighth,\cite{Knak} and the method of Abragam
was generalized in order to account for the values of these higher
moments.\cite{Par1,Par2} Experimentally, the fid curves of the
standard NMR-system CaF$_2$ were determined with very high precision
by Engelsberg and Lowe.\cite{Engel} Important progress in determining
the response function from the equations of motion was made by Becker
{\it et al.}\cite{Becker} and by Lundin.\cite{Lun1,Lun2} The method of
Lundin, which has recently been developed further by
Shakhmuratov,\cite{Shak} is reminiscent of the simple physical
interpretation made by Lefmann and collaborators of the double peak
structure in $^{13}$C diamond, where the splitting is considered to be
due to the strongly-coupled pairs of nearest neighbors, and the width to a
Gaussian-like distributed random-field from other
neighbors.\cite{Lef,Schaum}

In the present work the equations of motion of the Green function are
analyzed. One main result is that a solution of these equations can be
found when only the $z$-components of the spins are coupled. The Green
function in this case is determined by an infinite continued-fraction,
which may be solved accurately by numerical methods. This result is
then generalized to include the effects of the $x$- and
$y$-components of the dipole--dipole interaction. In this step, we
utilize the analytic results obtained previously for the higher
moments. A numerical analysis is then carried through in the cases
corresponding to CaF$_2$ and $^{13}$C diamond. The results
obtained by the present method agree closely with the experimental
results obtained for CaF$_2$. In the case of  $^{13}$C diamond there are
discrepancies, which are probably mostly of
experimental origin.

\section{Equations of motion}
\label{sec:level2}
We want to consider the linear magnetic response of a lattice of nuclear
moments subjected to a large stationary field. The normal situation in
the case of a NMR experiment is that $kT$ is much larger than the
mean-field splitting between the nuclear levels, which is again much
larger than the energy shifts due to the coupling between the nuclear
moments. These conditions we shall assume to be satisfied throughout.
We shall only treat the case of $I=1/2$ nuclei placed in a non-metallic
crystal. In this case we need only to consider the classical coupling
of the nuclear moments and, neglecting non-linear effects, the
Hamiltonian may effectively be reduced to\cite{Abra,Slichter}
\begin{equation}
{\cal H}=-{\Delta\over2}\sum_i\sigma_i^z-\sum_{ij} J_x^{}(ij)s_i^+s_j^-
-{1\over4}\sum_{ij} J_z^{}(ij)\sigma_i^z\sigma_j^z\,.\label{5}
\end{equation}
$\Delta=g\mu_N^{}H_z^{}$ is the Zeeman splitting between the
$I_z^{}=\pm {1\over2}$ states, $g$ is the nuclear $g$-factor, and the
gyromagnetic ratio $\gamma=g\mu_N^{}/\hbar$. The operators
$s_j^\pm=I_{xj}^{}\pm iI_{yj}^{}$ and $\sigma_j^z=2I_{zj}^{}$. An
effective dipole-coupling parameter is defined as
\begin{equation}
D_{ij}^{}={{3(1-3\cos^2\theta_{ij}^{})}\over{4R_{ij}^3}}\,(g\mu_N^{})^2\,,
\label{3}
\end{equation}
where $R_{ij}$ is the distance between the $i$th and $j$th nucleus, and
$\theta_{ij}^{}$ is the angle between the line connecting the two
nuclei and the $z$-axis. The contribution of the dipole--dipole
interaction is then included in Eq.\ (\ref{5}) via
\begin{equation}
J_x^{}(ij)={\textstyle{1\over3}}D_{ij}^{}\quad;\quad
J_z^{}(ij)=-{\textstyle{2\over3}}D_{ij}^{}\,.\label{4}
\end{equation}

The frequency-dependent Green functions are defined as
\begin{equation}
G(\hat{A})=\lim_{\eta\to0^+}\int_{-\infty}^{\infty}
(-{\textstyle{i\over\hbar}})\theta(t)\langle[\hat{A}(t),
{\textstyle\sum_j} s_j^-]\rangle
e^{i(\omega+i\eta)t}dt\,,\label{6}
\end{equation}
where $\theta(t)$ is the step function, and the
equation-of-motion\cite{Zub,JJ} is
\begin{equation}
\hbar\omega G(\hat{A})-G([\hat{A},{\cal H}])=
\langle[\hat{A},{\textstyle\sum_j}s_j^-]\rangle\,. \label{7}
\end{equation}
Due to our general assumptions, we can neglect the dipole coupling
in the determination of the thermal average on the r.h.s.\ of
this equation, and only $\langle[s_i^+,\sum_j s_j^-]\rangle=\langle
\sigma_i^z \rangle\simeq \Delta/kT$ is non-zero to leading order in
$1/kT$. The linear response observed in NMR experiments is
proportional to the imaginary part of the Green function $G(s_i^+)$.
Utilizing the special properties of the spin-1/2 operators, e.g.\
$(\sigma_i^z)^2=1$,  Eqs.\ (\ref{5}) and (\ref{7}) lead to the
following equation:
\begin{equation}
\varepsilon G(s_i^+)=- \langle\sigma_i^z\rangle
+\sum_j[J_x^{}(ij)G(\sigma_i^zs_j^+)-J_z^{}(ij)G(s_i^+\sigma_j^z)]
\label{8}
\end{equation}
with $\varepsilon=\Delta -\hbar\omega$.
The next-order Green functions are determined by ($i\neq j$)
\begin{eqnarray}
\varepsilon G(s_i^+\sigma_j^z&&)=J_x^{}(ij)G(s_j^+)-J_z^{}(ij)G(s_i^+)
\nonumber\\
{}+\sum_k{}'&&\Big[J_x^{}(ik)G(\sigma_i^z\sigma_j^zs_k^+)-J_z^{}(ik)
G(s_i^+\sigma_j^z\sigma_k^z)\nonumber\\
&&{}+2J_x^{}(jk)\{G(s_i^+s_j^+s_k^-)-G(s_i^+s_j^-s_k^+)\}\Big]\,,
\label{9}
\end{eqnarray}
where the primed sum means that the summation index ($k$) has to be
different from the other indices ($i$ or $j$). Next we find, when all
indices $i$, $j$, and $k$ are different from each other,
\begin{eqnarray}
\varepsilon G(s_i^+\sigma_j^z\sigma_k^z)&&{}=
J_x^{}(ik)G(\sigma_j^zs_k^+)+J_x^{}(ij)G(s_j^+\sigma_k^z)\nonumber\\
&&{}-J_z^{}(ik)G(s_i^+\sigma_j^z)-J_z^{}(ij)G(s_i^+\sigma_k^z)\nonumber\\
{}+\sum_l{}'\Big[J_x^{}(i&&l)G(\sigma_i^z\sigma_j^z\sigma_k^zs_l^+)
-J_z^{}(il)G(s_i^+\sigma_j^z\sigma_k^z\sigma_l^z)\nonumber\\
{}+2J_x^{}&&(jl)\{G(s_i^+s_j^+\sigma_k^zs_l^-)-
G(s_i^+s_j^-\sigma_k^zs_l^+)\}\nonumber\\
{}+2J_x^{}&&(kl)\{G(s_i^+\sigma_j^zs_k^+s_l^-)-
G(s_i^+\sigma_j^zs_k^-s_l^+)\}\Big]\label{10}
\end{eqnarray}
and
\begin{eqnarray}
2\varepsilon G(s_i^+s_j^+s_k^-)&&{}=J_x^{}(ik)\{G(\sigma_i^zs_j^+)-
G(s_j^z\sigma_k^z)\}\nonumber\\
&&{}+J_x^{}(jk)\{G(s_i^+\sigma_j^z)-G(s_i^+\sigma_k^z)\}\nonumber\\
{}+2\sum_l{}'\Big[J_x^{}(i&&l)G(\sigma_i^zs_j^+s_k^-s_l^+)\nonumber\\
{}+J_x^{}(jl)G(&&s_i^+\sigma_j^zs_k^-s_l^+)
-J_x(kl)G(s_i^+s_j^+\sigma_k^zs_l^-)\nonumber\\
{}-\{J_z^{}(i&&l)+J_z^{}(jl)-J_z^{}(kl)\}G(s_i^+s_j^+s_k^-
\sigma_l^z)\Big]\,.\label{11}
\end{eqnarray}

The major complications in the equations of motion above are due to
$J_x^{}(ij)$. The Green function $G_z^{}(s_i^+)$, determined by these
equations when $J_x^{}(ij)\equiv0$, only depends on higher-order Green
functions $G_z^{}(s_i^+\sigma_j^z\sigma_k^z\cdots)$ involving one
$s_{}^+$ operator, and in the next order:
\begin{eqnarray}
\varepsilon &&G_z^{}(s_i^+\sigma_j^z\sigma_k^z\sigma_l^z)=
-J_z^{}(il)G_z^{}(s_i^+\sigma_j^z\sigma_k^z)\nonumber\\
&&{}-J_z^{}(ik)G_z^{}(s_i^+\sigma_j^z\sigma_l^z)
-J_z^{}(ij)G_z^{}(s_i^+\sigma_k^z\sigma_l^z)\nonumber\\
&&{}-\sum_m{}'J_z(im)G_z^{}(s_i^+\sigma_j^z\sigma_k^z\sigma_l^z\sigma_m^z)\,,
\label{12}
\end{eqnarray}
where the first three terms all contribute the same to
$G_z^{}(s_i^+)$ and thus may effectively be replaced by
$-3J_z^{}(il)G_z^{}(s_i^+\sigma_j^z\sigma_k^z)$. Similarly, the two
terms in front of the summation sign of Eq.\ (\ref{10})
may be replaced by $-2J_z^{}(ik)G_z^{}(s_i^+\sigma_j^z)$. Hence the
hierarchy of Green functions determining $G_z^{}(s_i^+)$ has a
transparent structure. Introducing a vector with the components
$G_z^{}(s_i^+)$, $J_z^{}(ij)G_z^{}(s_i^+\sigma_j^z)$,
$J_z^{}(ij)J_z^{}(ik)G_z^{}(s_i^+\sigma_j^z\sigma_k^z)$,
etc.\ the equations of motion may be written as an infinite-dimensional
matrix equation, which has the solution of an infinite continued-fraction
\begin{equation}
G_z(s_i^z)={\quad{}-\langle \sigma_i^z\rangle\hfill\over{\varepsilon
-\displaystyle\sum_j{\strut\quad J_z^2(ij)\hfill\over{\varepsilon
-\displaystyle\sum_k{}^\prime{\strut\quad  2J_z^2(ik)\hfill\over{\varepsilon
-\displaystyle\sum_l{}^\prime{\strut 3J_z^2(il)\over{
\varepsilon-Q_4^z(ijkl)}}}}}}}}\label{13}
\end{equation}
with
\begin{equation}
Q_4^{z}(ijkl)=\sum_m{}'{{4J_z^2(im)}\over{\varepsilon-Q_5^z(ijklm)}}
\label{14}
\end{equation}
etc. Notice that $Q_4^{z}(ijkl)$ depends implicitly on the indices $j$,
$k$, and $l$, because the summation index $m$ has to be different from
all the preceding ones, as indicated by the prime.

The conjecture is now that the $J_x^{}(ij)$ coupling may be included as
a perturbation, so that the result may still be written
like the infinite continued-fraction above, except that the dependence
on the coupling parameters is altered, i.e.\ $Q_p^{z}$ in (\ref{13}) is
replaced by
\begin{equation}
Q_p^{}(ij\cdots)=\sum{}'{X_p^{}(ij\cdots)\over\varepsilon
-Q_{p+1}^{}(ij\cdots)}\,,\label{15}
\end{equation}
where the parameters $X_p^{}(ij\cdots)$ are determined by the
equations of motion when the $J_x^{}(ij)$ coupling is included.
Expanding the infinite continued-fraction in powers of
$1/\varepsilon$, we get
\begin{eqnarray}
G(&&s_i^+){}= -{\langle\sigma_i^z\rangle\over\varepsilon}\Big[
1+{1\over\varepsilon^2}\sum_j{}'X_1^{}(ij)+{1\over\varepsilon^4}
\sum_j{}'X_1^{}(ij)\nonumber\\&&{}\times
\big\{\sum_{j'}{}'X_1^{}(ij')+\sum_k{}'X_2^{}(ijk)\big\}
+ \cdots\Big]\,.\label{16}
\end{eqnarray}
Assuming all nuclei to be placed in equivalent surroundings, i.e.\
$G(s_i^+)=G(s_j^+)$, then we get from Eqs.\ (\ref{8})--(\ref{11})
\begin{eqnarray}
&&G(s_i^+){}= -{\langle\sigma_i^z\rangle\over\varepsilon}\Big[
1+{1\over\varepsilon^2}\sum_j{}D_{ij}^2
+{1\over\varepsilon^4}\sum_jD_{ij}^2\label{17}\\&&{}\times
\big\{\sum_{j'}D_{ij'}^2+{1\over9}\sum_k{}'
(7D_{ik}^2+5D_{jk}^2+6D_{ik}^{}D_{kj}^{})\big\}
+ \cdots\Big]\nonumber
\end{eqnarray}
utilizing that the summation indices may be interchanged (the prime on
the $j$-sums is unnecessary as $D_{ii}^{}=0$). A comparison of
(\ref{16}) and (\ref{17}) then allows the identification of $X_1^{}$
and $X_2^{}$. The distinction between $D_{ik}^{2}$ and $D_{jk}^{2}$ in
the last sum is arbitrary, and because $G(s_i^+)=(1/N)\sum_i G(s_i^+)$
the result should be same if the sum over $j$, in the first step of
the infinite continued-fraction, is replaced by a sum over $i$.
In order to ensure this, also when introducing the higher-order terms,
it is sufficient to assume the solution to be symmetric in $i$ and $j$:
\begin{mathletters}
\begin{eqnarray}
X_1^{}(ij)&=&D_{ij}^2\label{18}\\
X_2^{}(ijk)&=&{2\over3}\big\{D_{ik}^2+D_{jk}^2+
D_{ik}^{}D_{kj}^{}\big\}\,.
\label{19}
\end{eqnarray}
\end{mathletters}
The moments of the spectral energy-distribution, defined with respect
to the mean value $\Delta$, are
\begin{equation}
M_n^{}=\langle\varepsilon_{}^n\rangle=\int_{-\infty}^\infty\varepsilon_{}^n
G''(s_i^+;\varepsilon)d\varepsilon\Big/\int_{-\infty}^\infty\
G''(s_i^+;\varepsilon)d\varepsilon\,,\label{20}
\end{equation}
where $G''(s_i^+;\varepsilon)$ denotes the imaginary part of the Green
function ($M_n^{}$ is $(2\pi\hbar)^n$ times the usual frequency
moments). The odd moments vanish by symmetry, and an elementary
result is
\begin{equation}
G(s_i^+)= -{\langle\sigma_i^z\rangle\over\varepsilon}\Big[1+
{1\over\varepsilon^2}M_2^{}+{1\over\varepsilon^4}M_4^{}+\cdots\Big]\,,
\label{21}
\end{equation}
which shows that (\ref{17}) is equivalent to an expansion of
$G(s_i^+)$ in terms of the spectral moments. Introducing the notation
$S_n^{}=\sum_jD_{ij}^n$, the moments are
\begin{mathletters}
\begin{eqnarray}
M_2^{}&=&S_2\label{22}\\
M_4^{}&=&{1\over3}\big\{7S_2^2-4S_4^{}+2S_2^{}
\sum_kD_{ik}^{}D_{kj}^{}\big\}
\label{23}
\end{eqnarray}
\end{mathletters}
in accordance with previous results.\cite{VV,Knak}

The complexity in performing the expansion (\ref{17}) of $G(s_i^+)$
increases dramatically in the next order. Therefore we make use of the
additional assumption that terms which involve products of two
different $D$-couplings can be neglected in $X_p^{}$, when $p\geq3$.
The coupling $D_{ij}^{}$ takes on both positive and negative values
and $D_{ij}^{}$ integrated over a sphere vanishes, whereas in
contrast $D_{ij}^2$ is always positive. Advancing one more step in
the hierarchy of Green functions, beyond that of Eqs.\ (\ref{10}) and
(\ref{11}), and keeping track of only the squared coupling terms, we
find the linear combination
$X_3^{}(ijkl)=aD_{il}^2+bD_{jl}^2+cD_{kl}^2$ to be determined by
$a+b+c=41/18$, and further that a solution symmetric in $i$ and
$j$ ($a=b$) implies $c=0$, thus
\begin{equation}
X_3^{}(ijkl)={41\over36}\big\{D_{il}^2+D_{jl}^2\big\}\,.\label{24}
\end{equation}
This result predicts the sixth moment $M_6^{}$ to be
$$
{1\over27}\big\{229S_2^3
-373S_2^{}S_4^{}-171S_6^{}+41\sum_{jk}{}
D_{ij}^2D_{ik}^2D_{kj}^2\big\}
$$
plus the terms deriving from $D_{ik}^{}D_{kj}^{}$ in $X_2^{}(ijk)$,
which is consistent with the analytical result for $M_6^{}$ derived by
Knak Jensen and Kj\ae rsg\aa rd Hansen.\cite{Knak} It is possible to
include most of the terms in $M_6^{}$ depending on the linear factors,
but the price is that $Q_2^{}(ij)$ has to be divided into at least
two continued fractions, which makes the procedure somewhat arbitrary.
More importantly, in principle, is the occurrence\cite{Knak} of the term
$-(8/27)\sum{}
D_{ij}^{}D_{ik}^{}D_{il}^{}D_{lk}^{}D_{kj}^{}D_{jl}^{}$ in $M_6^{}$.
It falls outside the present scheme and cannot be incorporated into
the continued fraction in any simple way, thus indicating a limit to the
present procedure. However, as discussed above, the contributions
to $M_6^{}$ which are neglected in Eq.\ (\ref{24}) are
expected to be small, and this is supported by the numerical analysis
discussed below.

In the higher order it is not important to discriminate between the
different quadratic contributions, e.g.\ between $D_{im}^2$, $D_{jm}^2$,
$\cdots$ in $X_4^{}(ijklm)$. In any case, such a separation plays no
role in the following numerical analysis, and we may assume,
\begin{equation}
X_p^{}(ijk\cdots\gamma)={\alpha_p^{}\over2}\big\{D_{i\gamma}^2
+D_{j\gamma}^2\big\} \quad;\quad p\geq4 \label{26}
\end{equation}
corresponding to Eq.\ (\ref{24}).
Using the result of Knak Jensen and Kj\ae rsg\aa rd Hansen\cite{Knak}
that the leading order term in $M_8^{}$ is $(11031/243) S_2^4$, we find
$\alpha_4^{}=2819/738$. Introducing an effective $\alpha_p^{}$
for the lower-order terms also, we get the following sequence of values for
$\alpha_p^{}$:
$$
1,\quad {4\over3},\quad{41\over18},\quad{2819\over738}
$$
for $p=1$, 2, 3, and 4. The final Green function is not much dependent
on how the series is continued, except that $\alpha_p^{}$ increases
with $p$ (which is guarantied by the $J_z^{}(ij)$-coupling). Although
the four moments $M_2^{}$--$M_8^{}$ do not determine the response,
they put strong limits on the kind of variation which may be achieved
by varying the higher-order terms. However, our ignorance of how the
$\alpha_p^{}$ series continues leaves us a single degree of freedom,
which we have utilized for a small over-all adjustment of the calculated
response functions, using for this purpose the accurate
experimental results for CaF$_2$. We have tried to use
a linear extrapolation (with a slope between 1.15 and 1.2), but
the $\alpha_p^{}$-values above are described closely by $\alpha_p^{}=
1+0.033(p-1)+0.302(p-1)^2$, i.e.\ $\alpha_p^{}$ seems to increase
quadratically rather than linearly with $p$, and we obtained a slightly
better result by using the following quadratic extrapolation:
\begin{equation}
\alpha_p^{}=1+0.327(p-1)^2\quad;\quad p\geq5\,.\label{27}
\end{equation}

In the numerical evaluation of the Green function we have to make some
additional approximations. The infinite continued-fraction
\begin{equation}
G(s_i^z)={\quad{}-\langle \sigma_i^z\rangle\hfill\over{\varepsilon
-\displaystyle\sum_j{\strut\quad D_{ij}^2\hfill\over{\varepsilon
-Q_2^{}(ij)}}}}\label{29}
\end{equation}
becomes less and less dependent on the actual values of
$Q_p^{}(ij\cdots)$ the larger $p$ is, and to a first approximation
we assume that the summation index only needs to be different from
$i$ and $j$. In this approximation
$\sum{}'X_p^{}(ijk\cdots\gamma)\simeq\alpha_p^{}V(ij)$ with $V(ij)=
S_2^{}-D_{ij}^2$ for $p\geq3$, hence
\begin{equation}
Q_2^{}(ij)\simeq{{4\over3}V(ij)+{2\over3}\sum_k^{}D_{ik}^{}D_{kj}^{}
\hfill\over{\varepsilon
-\displaystyle{\strut\alpha_3^{}V(ij)\hfill\over
{\varepsilon
-\displaystyle{\strut \alpha_4^{}V(ij)\hfill\over{\varepsilon
-\displaystyle{\strut \ddots}}}}}}}\,.\label{30}
\end{equation}
The infinite continued-fraction has to be determined for each
value of the summation index $j$. This is done by assuming
$\alpha_p^{}$ to be constant for $p\geq n$, i.e.\
$Q_n^{}(ij)\simeq\alpha_n^{}V(ij)/\{\varepsilon-Q_n(ij)\}$
or
$$
Q_n^{}(ij)={1\over2}\Big\{\varepsilon-
\sqrt{\varepsilon^2-\alpha_n^{}V(ij)}\,\Big\}\,.
$$
The minus sign in front of the square root is the only choice which leads to
the right sign of $G''(s_i^+)$. Even for rather large values of $n$,
this termination of the infinite continued-fraction gives rise to spurious
oscillations in the calculated response function (as function of $n$
or $\varepsilon$), which are, however, found to cancel out in the mean
value of the response function calculated for two successive values of
$n$. With the use of this averaging procedure it is found that the
result becomes independent of $n$, when $n$ is larger than about
20--30 (we have used $n=50$ and 51 in the final calculations). The
approximation made in Eq.\ (\ref{30}) for $Q_2^{}(ij)$ may be improved,
and to next order we get
\begin{equation}
Q_2^{}(ij)={2\over3}\sum_k{}'
{D_{ik}^2+D_{jk}^2+ D_{ik}^{}D_{kj}^{}
\hfill\over{\varepsilon
-\displaystyle{\strut\alpha_3^{}W(ijk)\hfill\over
{\varepsilon
-\displaystyle{\strut \alpha_4^{}W(ijk)\hfill\over{\varepsilon
-\displaystyle{\strut \ddots}}}}}}}\label{31}
\end{equation}
with $W(ijk)=S_2^{}-D_{ij}^2-(D_{ik}^2+D_{jk}^2)/2$. The effective
coupling parameter $D_{ij}^2$ is proportional to $R_{ij}^{-6}$, but
although this coupling decreases much faster than the dipole coupling
itself, the $j$ sum has to be extended over 10--20,000 neighbors in
order to obtain an acceptable accuracy. Fortunately, the modifications
introduced by the replacement of Eq.\ (\ref{30}) by Eq.\ (\ref{31})
only influence the final response function weakly, and it is only
necessary to apply the improved expression for $Q_2(ij)$ when $j$ is
one of the 20--30 neighbors which are coupled most strongly to the
$i$th site. The result (\ref{31}) may be considered to correspond to
the analytical result derived above. The next step, which involves
taking into account that the summation indices in $Q_4(ijkl)$ have to be
different from $l$, presupposes a more detailed determination of
$X_4(ijklm)$ than given by Eq.\ (\ref{26}), and requires much more
extensive numerical calculations. Considering Eq.\ (\ref{31}) to be
the starting equation, we may say in short that all additional
approximations made in the numerical analysis are kept at such a level
that their influence on the final results is insignificant.

\section{Comparison with experiments}
\label{sec:level3}
\subsection{Calcium fluoride}
CaF$_2$ has for a long time served as a standard material for
comparisons between NMR line-shape theories and
experiments.\cite{Abra,Engel} The stable $^{19}$F isotope  in CaF$_2$
with $I=1/2$ and $g=5.25454$ ($\gamma=25166.2$ rad\,s$^{-1}$Oe$^{-1}$)
is positioned on a simple cubic lattice with the lattice parameter
$a/2=2.72325$ \AA\ in the low temperature limit.\cite{Engel} The
nuclear magnetic resonance in CaF$_2$ was measured by Bruce in
1957 with the field applied along the three high symmetry
directions,\cite{Bruce} and fourteen years later Engelsberg and Lowe
repeated the measurements and established with a high degree of
accuracy the fid curve in each of the three cases.\cite{Engel} We
have used their parametrized experimental results for calculating the
corresponding Fourier-transformed response functions which are shown
in Fig.\ \ref{fig1}. These results should benefit from the great
precision by which the fid curves were measured, and it has been
checked that the moments of the resonance curves (up to the eighth)
are those reported\cite{Engel} by Engelsberg and Lowe in their
Table IV. The results of Bruce,\cite{Bruce} in the cases where the
field is along $[100]$ or $[110]$, agree well with those derived from
the experimental fid curves of Engelsberg and Lowe, when the data are
corrected for the change of the lattice constant which occurs between
room temperature and 4.2 K. Bruce's result in the case of a field
along $[111]$ has a second moment which is about 13\% larger than the
theoretical value,\cite{VVrem} but a folding of the results of Engelsberg
and Lowe with a Gaussian with a width which corrects for the difference
between the second moments leads to a good coincidence of the two set of
experimental results in this case also.

The experimental results in Fig.\ \ref{fig1} are compared with the
predictions of Eq.\ (\ref{31}). The intensities shown in the
figure  are $-(g\mu_N^{}/\langle\sigma_i^z\rangle)G''(s_i^+)$ as
functions of the field parameter $\varepsilon/g\mu_N^{}$, and the
experimental results have been scaled so that the total
integrated intensity is equal to $\pi$ in all the cases shown. As may
be seen in the figure, the agreement between theory and experiment is
very good, but there are small systematic deviations, which are
at most 2--3\% of the intensity at zero frequency. Fig.\ \ref{fig2}
shows a blow-up of the intensity differences between the experimental
results and the theory in CaF$_2$. It is difficult to determine the
origin of these deviations, but they are most likely due to the
approximations made in the theory rather than to the experimental
uncertainties.

The corresponding spectral moments are given in Table \ref{table1}.
The experimental moments $M_2$ ($\widetilde{M_2}=M_2/g\mu_N^{}$ in
Table \ref{table1}) and $M_4$ are in good agreement with the
theoretical values. $M_6/M_2^3$ and $M_8/M_2^4$, which are determined
with experimental uncertainties of respectively about 7\% and 15\%,
are systematically smaller than predicted by the theory. However,
one should take into consideration that the exponentially decaying
tail, in for instance the [111] case above a field of 9.5 Oe,
accounts for 0.1\% of the intensity but contributes as much
as 11\% and 26\% to respectively the sixth and the eighth moment.
Actually, we may say that the parametrized experimental results of
Engelsberg and Lowe, including an extrapolation of the long-time
behavior of the fid curves, account for a surprisingly large portion of
the tails. Concerning the comparison between the moments predicted by the
present theory and the in principle exact values obtained by a direct
calculation of the moments, we remark firstly that, since the theory should
predict the right second and fourth moments, the close coincidence seen
for these moments just reflects the accuracy of the numerical
analysis. Secondly, we notice that the values of $M_6$ and $M_8$
predicted by the theory only differ from the exact values by a few per
cent, which shows that the terms discarded in the expression for
$X_3(ijk)$ or $X_4(ijkl)$, (\ref{24}) or (\ref{26}), only have minor
effects on these moments.

The free-induction-decay curve is defined as the (normalized)
time-dependent correlation function of the averaged $I_x$-component in
the rotating frame set up by the stationary field, which may be shown
to be\cite{Abra}
\begin{equation}
F(t)=\int_{-\infty}^\infty G''(s_i^+;\varepsilon)\cos(\varepsilon
t/\hbar)d\varepsilon\Big/\int_{-\infty}^\infty G''(s_i^+;\varepsilon)
d\varepsilon\,.
\label{32}
\end{equation}
The fid curve in the case of a field along $[100]$ predicted by the
theory is compared with the experimental (parametrized) fid curve
observed by Engelsberg and Lowe\cite{Engel} in Fig.\ \ref{fig3}.
The agreement is close for times less than 50 $\mu s$. At longer times
some discrepancies develop and the times at which the fid curve
becomes zero are about 5--6\% larger than observed experimentally.
The period of the oscillations is greater when the field is applied
along the other symmetry directions, and here the calculated positions
of the zeros agree within 3--4\% with the experimental values. The
long-time form of the fid curve is well described by the simpler
exponentially decaying function\cite{Engel}
$$
F(t)\approx A e^{-\alpha t}\cos(\beta t+c)\,,
$$
instead of the expression (\ref{1}) introduced by Abragam, and the
theoretical values of the parameters $\alpha$ and $\beta$ are compared
with experiment in Table \ref{table2}.

\subsection{$^{13}$C diamond}
The present work was initiated by the NMR-experiments of Lefmann {\it
et al.}\cite{Lef} on $^{13}$C diamond. However, the experimental
conditions were less favorable in this system and the results
therefore not of the same high accuracy as in CaF$_2$. The $I=1/2$
carbon nucleus $^{13}$C has $g= 1.40437$ ($\gamma=6726.1$
rad\,s$^{-1}$Oe$^{-1}$) and is placed in a diamond lattice with the lattice
parameter $a=3.5666$ \AA\ at room temperature. The theoretical
resonance curves  $-(2\pi\hbar/\langle\sigma_i^z\rangle)G''(s_i^+)$
as functions of frequency $\varepsilon/2\pi \hbar$, when the field
is applied along the three symmetry directions, are compared with the
experimental results in Fig.\ \ref{fig4}. It is immediately seen
that, although the qualitative behavior is similar, the experimental
widths of the resonance curves are considerably larger than the
theoretical ones. This is also clearly reflected in the
spectral moments given in Table \ref{table3} (here $\widetilde{M}_2=
M_2/2\pi\hbar$). The experimental second moments were calculated by
Lefmann {\it et al.}, and we also give the values of the experimental
fourth moments estimated from the resonance curves in Fig.\
\ref{fig4}. In addition to the theoretical second and fourth moments, which
are equal to the exact values, we also give the calculated values of the
sixth and eighth moments, which are expected to be correct within a
few per cent.

The experiment shows that the resonance line in the $[111]$ case
splits into two peaks, as also predicted by the theory. The calculated
splitting of the resonance line is 8.56 kHz, which is near to the
observed value. The important parameter, which is decisive for the
occurrence of the splitting, is the minimum value of the ratio
\begin{equation}
r(ij)={\textstyle{4\over3}}\big\{S_2^{}-D_{ij}^2+
{\textstyle{1\over2}\sum_k} D_{ik}^{}D_{kj}^{}\big\}\big/ S_2^{}\,.
\label{33}
\end{equation}
In $^{13}$C diamond in the $[111]$ case the minimum value is found to
be 0.43 for the nearest neighbors along the direction of the field.
The resonance in CaF$_2$ when the field is along $[100]$ is just
on the threshold where it splits into two peaks, and here the minimum
value is 0.94, slightly smaller than 1, valid for the four nearest
neighbors in the plane perpendicular to the field. In any of the other
cases considered in the two systems the minimum ratio is larger than 1.

A comparison of the tabulated moments shows that the ratio between the
second and the higher moments in  CaF$_2$, when the field is along
$[111]$ and $[110]$, are nearly the same as in $^{13}$C diamond when
the field is along, respectively, $[100]$ and $[110]$. This
coincidence is also found to occur for the calculated line-shape
curves (most pronouncedly in the first of the two cases), if the
square root of the second moments are used as scale parameters.

The comparison between theory and experiment in $^{13}$C diamond shows
that the experimental NMR line shapes are distorted. The only additional
effects which may influence the results are the presence of impurities,
or the repetition rate of 3--4 spectra per minute used in the
experiment, which is relatively high compared with the long
spin--lattice relaxation time of 14--16 s.
To a first approximation the impurities give rise to an extra
Gaussian broadening of the spectra. This is concordant with the results
obtained in the $[100]$ case, which correspond to the theoretical curve
folded with a Gaussian with $\sigma=1.6$ kHz, but the differences
in the two other cases are not describable in this manner, which
indicates that the high repetition rate might have had some influence
on the results.

\section{Discussion and conclusion}
\label{sec:level5}
The NMR line shape due to the dipole coupling of $I=1/2$ nuclei has been
calculated approximately at high temperatures and high fields. The
result is determined in terms of an infinite continued-fraction, which
is relatively easy to handle by numerical methods. The spectra
derived by the present theory has the right second and fourth moments,
and the sixth and eighth moments are found to be close to the
values\cite{Knak} calculated directly in the case of a simple cubic system.
The theoretical results agree satisfactorily with the precise experimental
results\cite{Engel} of Engelsberg and Lowe for the fid curves in
CaF$_2$ obtained with the field along each of the three high symmetry
directions.

The present method bears some resemblance with the procedure of Parker
and Lado\cite{Par1,Par2}, who utilized the moment expansion directly
for a calculation of the line-shape curves. The results obtained by
the two methods are also quite similar but, considering the behavior
of the fid curves at long times, Table \ref{table2}, the lengths of
the periods predicted by the present theory are closer to the
experimental values than those obtained by Parker and Lado. The zeros of
the $[100]$ fid-curve in CaF$_2$ predicted by the approximate solution
derived by Becker {\it et al.}\ of the equations of motion agree even
more closely with experiment.\cite{Becker}  On the other hand their
values for the moments, beyond the second, are systematically
smaller than the correct values, which is probably mostly due to their
neglect of terms involving more than one $s^+$ operator like the
Green function $G(s_i^+s_j^+s_k^-)$ in Eq.\ (\ref{11}). The method of
Lundin\cite{Lun1,Lun2} is based on a small number of assumptions which
appear to be nearly fulfilled, and the improvements introduced by
Shakhmuratov\cite{Shak} lead to an excellent description of the
$[100]$ fid-curve in CaF$_2$.

The infinite-continued-fraction solution derived in the present work
may be improved in various ways. The most obvious one would be to
include some of those terms neglected in $X_3$ and $X_4$. We have
made some effort in this direction, but the improvements obtained
were small and not commensurate with the extra complications
appearing in the expression for the Green function and in the
numerical work. In any case they are more or less eliminated by the
arbitrariness connected to the extrapolation (\ref{27}) of $\alpha_p$
beyond $p=4$, which indicates that a further improvement in the theory
would have to include a determination of $\alpha_5$, in order to
reduce the importance of the assumption about the behavior of the
$\alpha_p$ series.

\begin{acknowledgements}
The author would like to thank K. Lefmann and collaborators for
providing him with the results of their measurements on $^{13}$C
diamond before publication. Useful discussions with K. Lefmann, F.
Berg Rasmussen and A.R. Mackintosh are gratefully acknowledged.
\end{acknowledgements}

\begin{table}
\caption{The spectral moments in CaF$_2$. The experimental values are
derived from the parametrized fid curves of Engelsberg and Lowe,$^{10}$
and the exact values of $M_6$ and $M_8$ are the results of Knak Jensen
and Kj\ae rsg\aa rd Hansen,$^8$ extrapolated to the infinite
lattice. $M_2$ is the same in each row of the table.}
\label{table1}
\begin{tabular}{llcccc}
&& $\widetilde{M}_2^{1/2}$ (Oe) & $M_4^{}/M_2^2$ & $M_6^{}/M_2^3$ &
$M_8^{}/M_2^4$ \\
\noalign{\vspace{1pt}}
\tableline
\noalign{\vspace{2pt}}
$[100]$ &
Expt. & 3.615\phantom{0} & 2.103\phantom{0} & 6.08\phantom{0} &
22.2\phantom{0} \\
&Theory & 3.6020 & 2.1244 & 6.427 & 26.57 \\
&Exact & 3.6021 & 2.1245 & 6.329 & 25.16 \\
\noalign{\vspace{2pt}}
$[110]$ &
Expt. & 2.191\phantom{0} & 2.236\phantom{0} & 6.93\phantom{0} &
26.7\phantom{0} \\
&Theory & 2.2174 & 2.3021 & 7.815 & 36.85 \\
&Exact & 2.2176 & 2.3022 & 7.709 & 34.96 \\
\noalign{\vspace{2pt}}
$[111]$ &
Expt. & 1.508\phantom{0} & 2.340\phantom{0} & 8.05\phantom{0} &
36.9\phantom{0} \\
&Theory & 1.4937 & 2.3693 & 8.540 & 43.78 \\
&Exact & 1.4940 & 2.3694 & 8.511 & 44.09 \\
\end{tabular}
\end{table}
\begin{table}
\caption{Parameters of the long-time form of the fid curves in CaF$_2$.
The experimental values are from Ref.\ [10].}
\label{table2}
\begin{tabular}{lcccc}
& $\alpha$ (calc.) & $\alpha$ (expt.) & $\pi/\beta$ (calc.) &
$\pi/\beta$ (expt.) \\
& ($\mu$s$^{-1}$) & ($\mu$s$^{-1}$) & ($\mu$s) & ($\mu$s) \\
\noalign{\vspace{1pt}}
\tableline
\noalign{\vspace{2pt}}
$[100]$ & 0.064 & 0.050 & 22.2 & 20.5 \\
\noalign{\vspace{1pt}}
$[110]$ & 0.043 & 0.041 & 32.8 & 30.6 \\
\noalign{\vspace{1pt}}
$[111]$ & 0.029 & 0.030 & 46.1 & 47.6 \\
\end{tabular}
\end{table}
\begin{table}
\caption{The spectral moments in $^{13}$C diamond. The experimental
  values are from Ref.\ [1].}
\label{table3}
\begin{tabular}{llcccc}
&& $\widetilde{M}_2^{1/2}$ (kHz) & $M_4^{}/M_2^2$ & $M_6^{}/M_2^3$ &
$M_8^{}/M_2^4$ \\
\noalign{\vspace{1pt}}
\tableline
\noalign{\vspace{2pt}}
$[100]$ &
Expt. & 2.1\phantom{000} & 3.7\phantom{000} & & \\
&Theory & 1.3710 & 2.3906 & 8.728 & 45.46 \\
\noalign{\vspace{2pt}}
$[110]$ &
Expt. & 4.5\phantom{000} & 2.5\phantom{000} & & \\
&Theory & 3.4294 & 2.2419 & 7.139 & 30.43 \\
\noalign{\vspace{2pt}}
$[111]$ &
Expt. & 5.0\phantom{000} & 2.0\phantom{000} & & \\
&Theory & 3.8800 & 1.7822 & 4.471 & 15.89 \\
\end{tabular}
\end{table}

\begin{figure}
\caption{The NMR line shape in CaF$_2$ when the field is along one of
the three high-symmetry directions, shown as a function of the field
parameter $\varepsilon/g\mu_N^{}$. The symbols indicate the
experimental results derived from the parametrized fid curves of
Engelsberg and Lowe,$^{10}$ and the solid lines are the theoretical
predictions.}
\label{fig1}
\end{figure}
\begin{figure}
\caption{The intensity difference in CaF$_2$ between the experimental
and theoretical results in Fig.\ 1.}
\label{fig2}
\end{figure}
\begin{figure}
\caption{The fid curve $F(t)$ in CaF$_2$ when the field is along [100].
The dashed lines are the experimental results of Engelsberg and
Lowe$^{10}$ and the solid lines are the theoretical results.}
\label{fig3}
\end{figure}
\begin{figure}
\caption{The NMR resonance line in $^{13}$C diamond as a function of
$\varepsilon/2\pi\hbar$ when the field is along $[100]$, $[110]$ and
$[111]$. The dashed lines show the experimental results of Lefmann {\it
  et al.}$^1$ and the solid lines are the theoretical curves.}
\label{fig4}
\end{figure}


\begin{references}
\bibitem{Lef} K. Lefmann, B. Buras, E. J. Pedersen, E. S. Shabanova,
P. A. Thorsen, F. B. Rasmussen and J. P. F. Sellschop, Phys.\ Rev.\  B
(1995, in press).
\bibitem{Pake} G. E. Pake, J.\ Chem.\ Phys.\ {\bf16}, 327 (1948).
\bibitem{Ped} B. Pedersen and D. F. Holcomb, J.\ Chem.\ Phys.\ {\bf38},
61 (1963).
\bibitem{Par1} G. W. Parker and F. Lado, Phys.\ Rev.\ B {\bf8}, 3081 (1973).
\bibitem{VV} J. H. Van Vleck, Phys.\ Rev.\ {\bf74}, 1168 (1948).
\bibitem{Bruce} C. R. Bruce, Phys.\ Rev.\ {\bf107}, 43 (1957).
\bibitem{Abra} A. Abragam, {\it The Principles of Nuclear Magnetism}
(Oxford University Press, London, 1961), Chap.\ 4.
\bibitem{Knak} S. J. Knak Jensen and E. Kj\ae rsg\aa rd Hansen, Phys.\
Rev.\ B {\bf7}, 2910 (1973).
\bibitem{Par2}G. W. Parker and F. Lado, Phys.\ Rev.\ B {\bf9}, 22 (1974).
\bibitem{Engel} M. Engelsberg and I. J. Lowe, Phys.\ Rev.\ B {\bf10},
822 (1974).
\bibitem{Becker} K.W. Becker, T. Plefka and G. Sauermann, J.\ Phys.\ C
{\bf9}, 4041 (1976).
\bibitem{Lun1} A. A. Lundin and A. V. Makarenko, Sov.\ Phys.\ JETP
{\bf60}, 570 (1984) [Zh.\ Eksp.\ Teor.\ Fiz.\ {\bf87}, 999 (1984)].
\bibitem{Lun2} A. A. Lundin,  Sov.\ Phys.\ JETP
{\bf75}, 187 (1992) [Zh.\ Eksp.\ Teor.\ Fiz.\ {\bf102}, 352 (1992)].
\bibitem{Shak} R. N. Shakhmuratov, J.\ Phys.\ Condens.\ Matt.\ {\bf3},
8683 (1991).
\bibitem{Schaum} K. Schaumburg, E. Shabanova and J. P. F. Sellschop,
J.\ Magn.\ Res.\  (Part A, Jan.\ 1995, in press).
\bibitem{Slichter} C. P. Slichter, {\it Principles of Magnetic
  Resonance}, 3rd edition (Springer-Verlag, New York, 1990).
\bibitem{Zub} D. N. Zubarev, Sov.\ Phys.\ Usp.\ {\bf3}, 320 (1960)
[Usp.\ Fiz.\ Nauk {\bf71}, 71 (1960)].
\bibitem{JJ} J. Jensen and A. R. Mackintosh, {\it Rare Earth
  Magnetism: Structures and Excitations} (Oxford University Press,
Oxford, 1991), Chap.\ 3.
\bibitem{VVrem} The theoretical values given by Bruce\cite{Bruce} are
obtained using a lattice parameter of $a/2=2.72$ \AA\ (about
0.4\% too small) and Eq.\ (13) in the paper of Van Vleck\cite{VV},
which can be replaced by the following more accurate version:
$\langle\Delta\nu^2\rangle_{\text{Av}}^{}
=37.326g^4\beta^4h^{-2}d^{-6}[{1\over3}
S(S+1)][(\lambda_1^4+\lambda_2^4+\lambda_3^4)-0.19483]$.
\end{references}
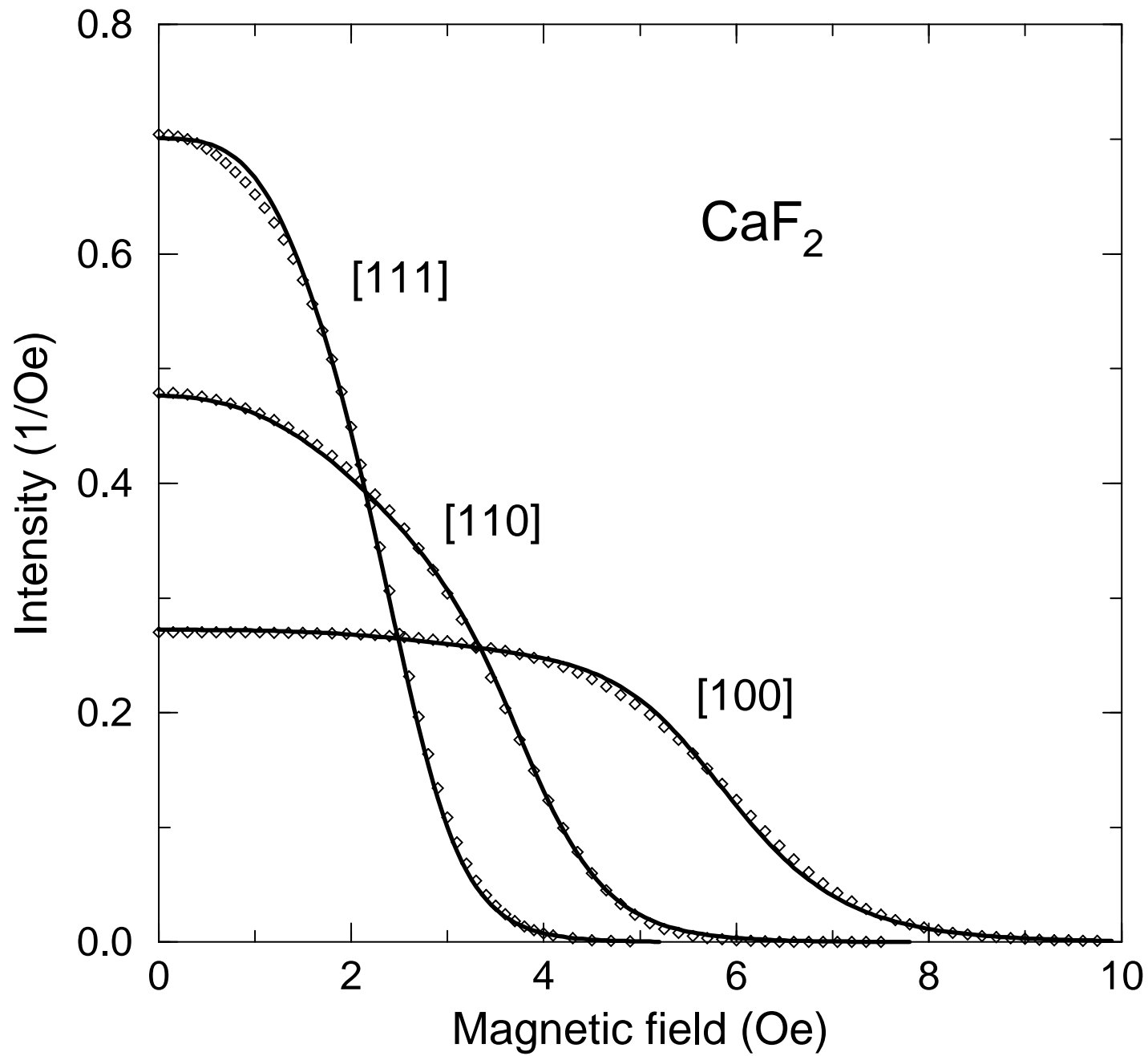
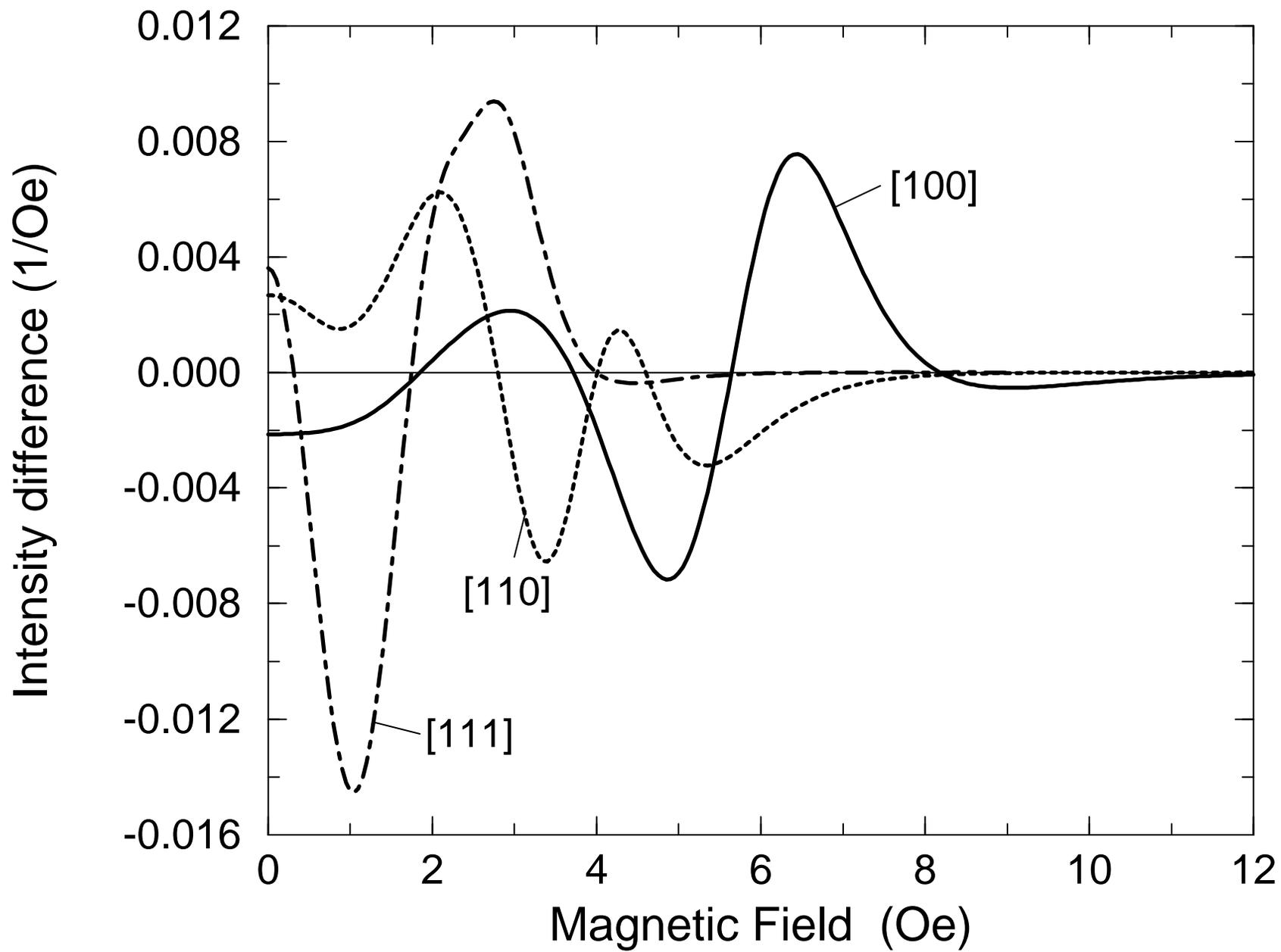
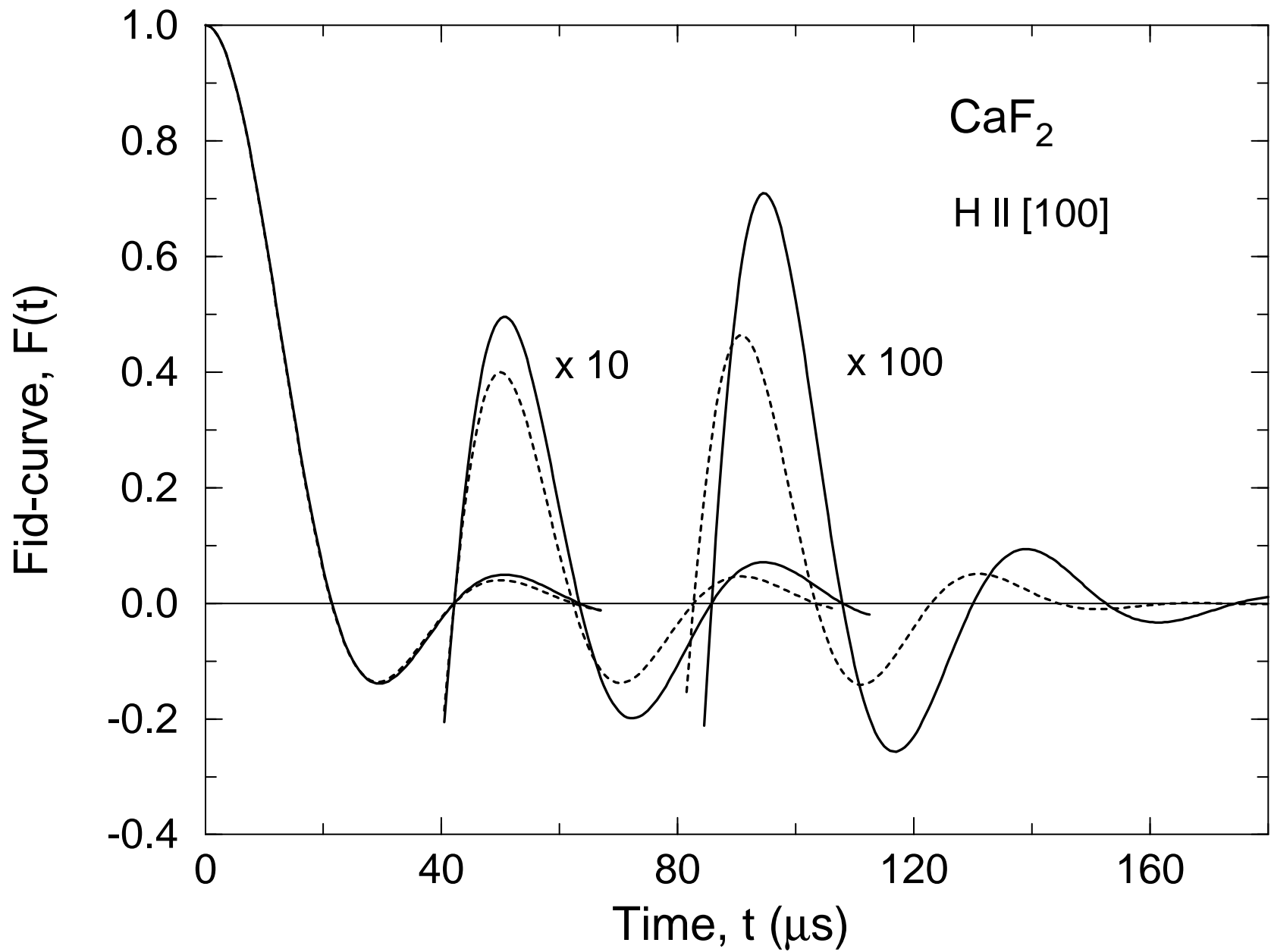
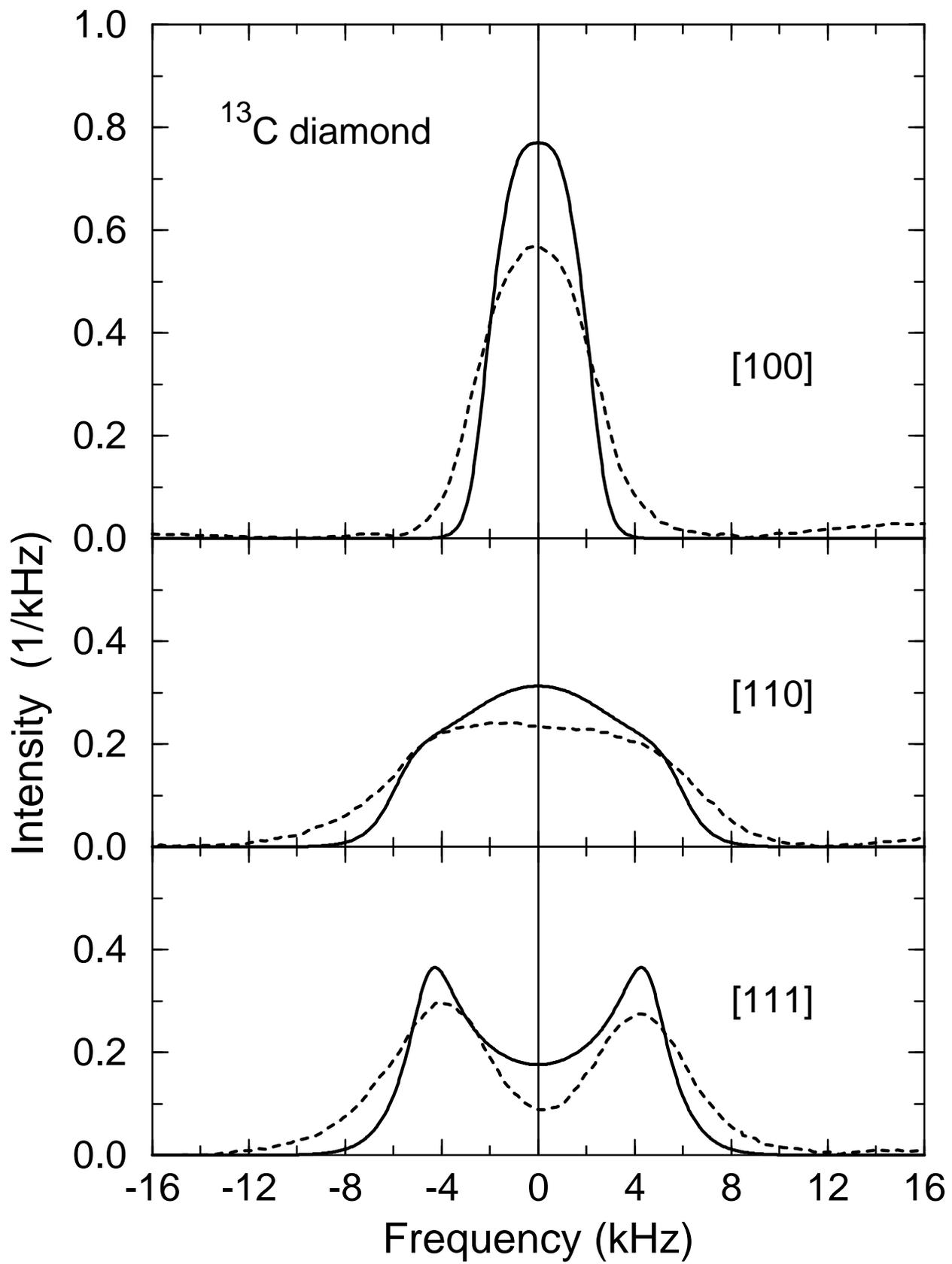
\end{document}